# SOLDERING OF MT-YBCO: METHOD TO PRODUCE SUPERCONDUCTIVE JUNCTIONS

Tatiana PRIKHNA[(1)], Wolfgang GAWALEK[(2)], Nikolay NOVIKOV[(1)], Viktor MOSHCHIL[(1)], Vladimir SVERDUN[(1)], Nina SERGIENKO[(1)], Alexey SURZHENKO[(2)], Michael WENDT[(2)], Tobias HABISREUTHER[(2)], Doris LITZKENDORF[(2)], Sergey DUB[(1)], Robert MÜLLER[(2)], Alexander KORDYUK[(3)], Silviya KRACUNOVSKA[(2)], Lydmila ALEXANDROVA[(1)]

[(1)]Institute for Superhard Materials of the Nat'l Ac. Sci of Ukraine, 2, Avtozavodskaya St., 04074, Kiev, Ukraine
[(2)]Institut für Physikalische Hochtechnologie e.V., Winzerlaer Str. 10, D-07743 Jena, Germany
[(3)]Institute of Metal Physics, 36, Vernadsky Pr., 03142, Kiev, Ukraine

A method of soldering of melt-textured $YBa_2Cu_3O_{7-\delta}$-based (MT-YBCO) superconductive blocks using a $TmBaCu_3O_{7-\delta}$ (Tm123) powder has been suggested. The method excludes stages of slow cooling during soldering and preparation of auxiliary layers. To estimate the critical current density ($j_c$) through the obtained junction, we have drilled rings from single-domained MT-YBCO blocks, cut into two pieces along the diameters and soldered using a Tm123 powder. The critical current density, $j_c$, (estimated using vibrating sample magnetometer) through the soldered ring was 34 kA/cm$^2$ in 0 T field at 77 K, which turned out to be even higher by a factor of approximately 1.5 than that through the single-domained material. The increase in $j_c$ through the soldered ring has been observed up to the 2.5 T field, while in the higher fields some decrease of $j_c$ took place. As a result of the field mapping of the trapped magnetic flux in soldered ring using a Hall probe, we obtained a regular truncated cone and the maximal level of trapped field for the ring 8 and 4 mm in outer and inner diameters was about 80 mT.

1. INTRODUCTION

Soldering of melt-textured $YBa_2Cu_3O_{7-\delta}$-based (MT-YBCO) superconductive blocks is a promising way of manufacturing large superconductive items of complex shapes, because up to now the existing technologies of melt-textured growth don't allow high quality parts larger than 40-60 mm to be produced. Due to the high level of superconductive characteristics such as critical current density ($j_c$) and levitation force, MT-YBCO can be successfully used in electromotors, flywheels, frictionless bearings, etc. The MT-YBCO structure is constituted from a single crystalline $YBa_2Cu_3O_{7-\delta}$ (or Y123) superconductive matrix (good samples usually contain one or a few single crystalline domains), in which fine inclusions of non-superconductive $Y_2BaCuO_5$ (or Y211) phase (in the amount of 20-30 %) are uniformly distributed. The single crystalline matrix allows high critical current density $j_c$ through the whole bulk because the weak links usually caused by the presence of grain boundaries are absent. The presence of fine inclusions of Y211 allows a further increase in the $j_c$ because of high density of dislocations, stacking faults, etc. around them, i.e. because of the high density of defect that are commensurable with the coherence length of Y123 superconductor and serves as pinning centers for magnetic flux. Melt-textured YBCO is manufactured using top-seeds from the composition with $Y_{1.5}Ba_2Cu_3O_x$ stoichiometry (with 1 wt% of $CeO_2$ addition). The matrix grows (crystallizes) in the temperature region from 1000 to 950 °C very slowly at a rate of about 0.3-1 K/h and during this process a growing Y123 crystal traps fine inclusions of Y211. $CeO_2$ is necessary for the refinement of Y211 grains.

The soldering is aimed at developing a binding layer between superconductive MT-YBCO blocks that were have similar superconductive and mechanical properties, which should be supported by the similar structure of the joining and the material. As a solder, a material with a lower melting temperature should be used. Good results have been obtained using as a solder a spacer-layer material cut from MT- TmBCO (melt-textured $TmBa_2Cu_3O_{7-\delta}$ with a 25 wt.% admixes of Y211, oxygenated 16 days at 450 °C in $O_2$) [1]. The soldering has been conducted in accordance with the "melt-textured" process (Fig.1, dashed line) that means holding for 6 h at 980 °C and a slow (0.5 K/h) cooling in the 980-955 °C region. Then the oxygenation at 440-480 °C in a separate process for 12 h - 2 weeks (depending on the sample size) took place[1]. By using the spacer layer material cut from MT-YBCO with Ag addition (that allows the melting temperature of MT-YBCO to be reduced) is also possible to obtain a good superconductive soldered seam[2]. Thus, the problem arises of manufacturing of large melt-textured high quality auxiliary layers to be used as solders. Besides, such a soldering process is complicated and rather time consuming.

We have developed a method to solder MT-YBCO using a $TmBaCu_3O_{7-\delta}$ (Tm123) powder that allows us to produce high-quality superconductive junctions between MT-YBCO parts excluding the stages of slow ("melt-textured") cooling during soldering and preparation of auxiliary layers. Using this method, we can solder a great quantity of parts differently arranged relative to each other. To estimate the critical current density $j_c$ through the soldered seam (methods proposed by H. Zheng et al.[3,4]), we used single-domained rings that were cut into two parts along the diameter and then soldered.

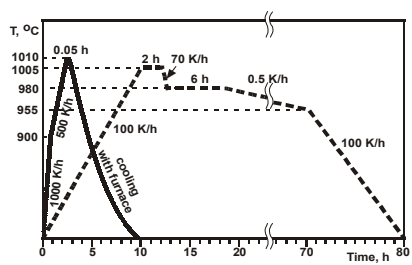

FIGURE 1
Regimes of soldering: solid line – used by the authors in the present study
dashed line – regime used by H. Zheng et al.[1]

2. EXPERIMENTAL

The rings (8 mm outer and 4 mm in inner diameters and 4-6 mm in height) have been drilled from single-domain MT-YBCO blocks. Rings were put into a vibrating sample magnetometer (VSM) and from magnetization loops the $j_c$ has been estimated using the equation:

$$j_c = 3M/[H \cdot \pi(R_2^3 - R_1^3)],$$

where M – magnetic moment, $R_1$, $R_2$ – inner and outer radii of the ring, respectively, H – the ring height.

The determined-from-DTA study temperature of the incongruent melting of MT-YBCO was 1016 °C and that of Tm123 was in the range from 984 to 986 °C.

Then these rings have been cut into two pieces along the diameters (the width of cut was about 0.7 mm) and soldered using a Tm123 powder (applied by sedimentation onto the soldered surfaces from the suspension in acetone) by regime shown on Fig.1 (solid line) with following oxygenation for 5.5 days. During soldering the parts to be joint were kept under 3 - 4 kg/cm$^2$ pressure using a special device. After soldering the ring has been again placed into the magnetometer to estimate the $j_c$. The maps of the trapped field in the ring have been investigated as well. To determine the flux-trapping capability, the samples were field cooled to 77 K in a 5 kG field oriented parallel to the c-axis (ring axis). The magnetic field was then turned off and the sample was transferred (without warming) to a liquid nitrogen Dewar equipped with a cryogenic scanning Hall probe. The Hall sensor has a active zone of about 50x50 µm. During the field scans, the probe was positioned at approximately 0.8 mm above the sample surface.

The structure of rings was studied using the SEM and polarizing microscopy. The magnetization hysteresis loops were obtained on an Oxford Instruments 3001 vibrating sample magnetometer (VSM). Hardness was measured on a Matsuzawa Mod. MXT-70 microhardness tester using a Vickers indenter. The bending strength was measured by the traditional method using a Universal Testing machine FM1000.

3. RESULTS AND DISCUSSION

In Fig.2 you can see the loops of magnetization of the starting single domain and soldered rings. Fig. 3 shows the $j_c$ calculated from magnetization loops. At 77 K - in 0 T field was $j_c$=34 kA/cm$^2$. The observed increase of the $j_c$ through the soldered ring by a factor of 1.5 as compared to the $j_c$ through the initial (uncut) ring can be explained by the fact that during the oxygenation that followed the soldering process the superconductive

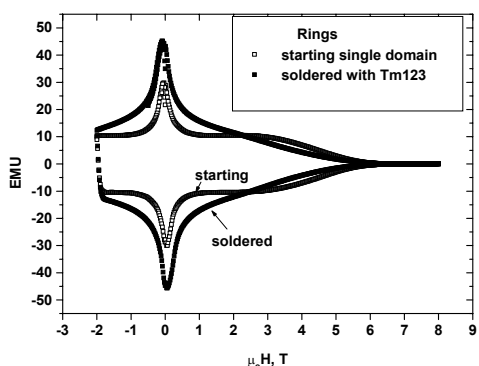

FIGURE 2

Magnetization loops of single domain (starting) and soldered (with Tm123 powder) MT-YBCO rings obtained by vibrating sample magnetometer.

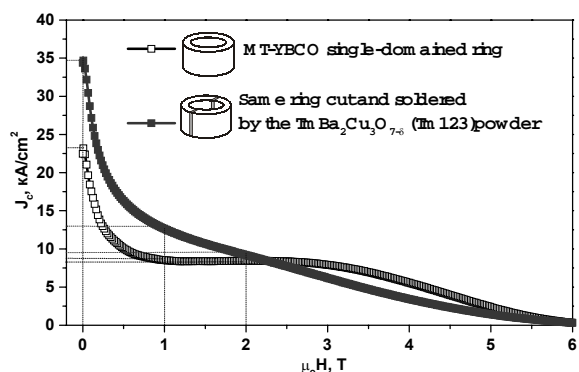

FIGURE 3

Critical current density, $j_c$, vs. magnetic field, $\mu_0 H$, of the single-domained ring and of the same ring cut and soldered by the Tm123 powder.

properties of the MT-YBCO material have been increased as well. The increase in the $j_c$ through the soldered ring have been observed up to the 2.5 T field, while in the higher fields some decrease of the $j_c$ took place.

The results of the field-mapping using Hall sensors have given the undoubted proofs that the high-quality junction have been obtained and are shown in Fig.4. We obtained a regular truncated cone and the maximal level of the trapped field for the ring was about 80 mT. The fact that the truncated cone is high by symmetric and that the upper surface has no valleys crevasse gives us undoubted proofs that the current through the seam is the same as through the material and the level of maximum trapped field indicates that the critical current density through the ring is higher than $10^4$ A/cm$^2$.

Fig. 5 demonstrates the structure of the seam and nearby joined material obtained using SEM and polarizing microscopy. One can see a good accommodation of the seam structure to the structure of MT-YBCO. In Fig.5 f the presence of Tm(Y)211 inclusions is observed. They have formed despite the fact that we haven't added any 211 phase (either Y211 or Tm123) to the Tm123 powder. In some places (Fig.5 c) it is well seen that even the structure of twins in the seam repeats the twin structure of the joined MT-YBCO. The width of the seam that can be estimated only from the COMP (composition) or Z-contrast image, obtained by SEM, turned out to be 50 μm. The Vickers microhardness of the seam and of the material was similar 4.57±

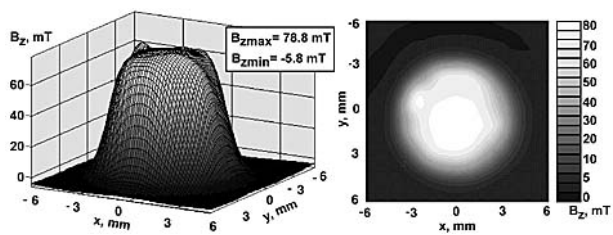

FIGURE 4

Trapped-field maps for the MT-YBCO ring soldered with Tm123.

0.82 GPa and 4.82±0.45 GPa (under 1.96-N load), respectively. When we estimated the bending strength of the seam, the breaking occurred mainly through the joined material, but not through the place of soldering and the strength was 28-32 MPa. This is the indicative of high mechanical properties of the junction.

4. CONCLUSIONS

Using a $TmBa_2Cu_3O_{7-\delta}$ (Tm123) powder as a solder we can obtain junctions between bulk superconductive parts of MT-YBCO, superconductive and mechanical properties of which are at the same level as those of the soldered material. The developed method allows us essentially simplify the soldering process and reduce the time of soldering. The calculated $j_c$ value through the best-soldered seam in 0 T field at 77 K was $j_c$=34 kA/cm$^2$. The observed increasing of $j_c$ through the soldered ring by a factor of 1.5 as compared to the $j_c$ through the initial (uncut) ring can be explained by the fact that during the oxygenation that followed the soldering process the superconductive properties of the MT-YBCO material have been increased as well. The increase in $j_c$ through the soldered ring have been observed up to the 2.5 T field. The results of field-mapping using Hall sensors gave us the undoubted proofs that the high-quality junction have been obtained. The structure of the seam practically repeats or slaves the structure of the soldered material.

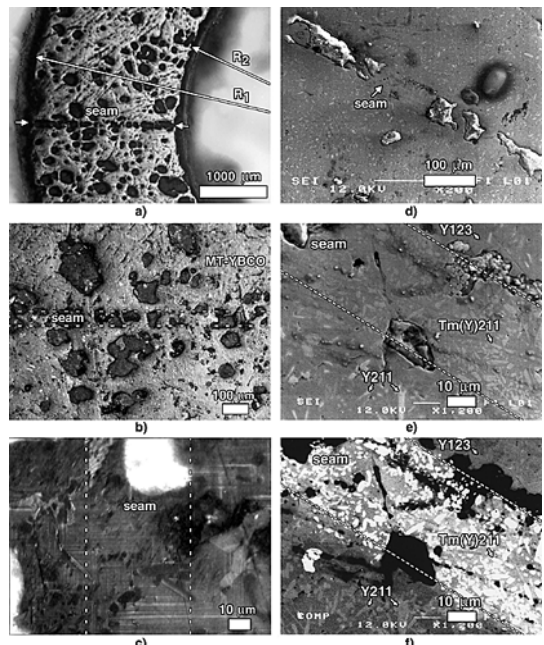

FIGURE 5

(a, b, c) the structure of the soldered seam obtained using polarizing microscopy at the different magnification

(d, e, f) SEM pictures of the seam soldered by Tm123: SEI –secondary electron image (d, e) and COMP - composition or Z- contrast image (f). In some cases to show the position of the seam the white dashed lines have been used.